\documentclass[11pt,twoside]{article}
\usepackage{asp2010}
\usepackage{natbib}

\resetcounters

\newcommand {\vect}[1]{\mbox{\boldmath $#1$}}

\newcommand {\inty}[2]{\int_{#1}^{#2}}

\newcommand {\pdif}[3][]{\frac{\partial^{#1}#2}{\partial#3^{#1}}}

\def\mart{\@ifnextchar[{\mart@@}{\mart@}}
\def\mart@@[#1]#2{\sqrt[#1]{\mathstrut{#2}}}
\def\mart@#1{\sqrt{\mathstrut{#1}}}

\newcommand {\Alfven}{Alfv\'{e}n}

\begin{document}

\title{Multi-Moment Advection scheme for Vlasov simulations}
\author{Takashi Minoshima$^1$, Yosuke Matsumoto$^2$, and Takanobu Amano$^3$
\affil{$^1$Institute for Research on Earth Evolution, Japan Agency for Marine-Earth Science and Technology, 3173-25, Syowa-machi, Kanazawaku, Yokohama 236-0001, Japan}
\affil{$^2$Department of Physics, Chiba University, 1-33, Yayoi-cho, Inage-ku, Chiba, 263-8522, Japan}
\affil{$^3$Department of Physics, Nagoya University, Furo-cho, Chikusa-ku, Nagoya 464-8602, Japan}}

\begin{abstract}
We present a new numerical scheme for solving the advection equation and its application to the Vlasov simulation.
The scheme treats not only point values of a profile but also its zeroth to second order piecewise moments as dependent variables, and advances them on the basis of their governing equations.
We have developed one- and two-dimensional schemes and show that they provide quite accurate solutions compared to other existing schemes with the same memory usage. 
The two-dimensional scheme can solve the solid body rotation problem of a gaussian profile with little numerical diffusion. This is a very important property for Vlasov simulations of magnetized plasma. 
The application of the scheme to the electromagnetic Vlasov simulation of collisionless shock waves is presented as a benchmark test.
\end{abstract}

\section{Introduction}\label{sec:introduction}
The kinematics of collisionless plasma has been studied in a wide variety of fields, such as in laboratory plasma physics, space physics, and astrophysics. Evolution of collisionless plasma and self-consistent electromagnetic fields is fully described by the Vlasov-Maxwell (or Vlasov-Poisson) equations. 
Thanks to recent development in computational technology, self-consistent numerical simulations of collisionless plasma have been successfully performed from the Vlasov-Maxwell system of equations.

One of numerical simulation methods for collisionless plasma is the so-called Vlasov simulation, in which the Vlasov equation is directly discretized on grid points in phase space. Compared to the most popular Particle-In-Cell (PIC) method \citep{PIC}, the Vlasov simulation is free from the statistical noise inherent to the PIC method. 
This advantage can allow us to study in detail such as wave-particle interaction, particle acceleration, and thermal transport processes, in which a high energy tail in the velocity distribution function plays an important role.
On the other hand, the Vlasov simulation requires a highly accurate scheme for the advection equation in multidimensions, to preserve characteristics of the Vlasov equation (i.e., the Liouville theorem) as much as possible.
It also requires larger computational cost than the PIC method.

A number of advection schemes have been proposed for the application to the Vlasov simulation thus far \citep[e.g.,][]{1976JCoPh..22..330C,1999CoPhC.120..122N,2001JCoPh.172..166F,2002JCoPh.179..495M}. Although the schemes have been succeeded especially in applying to the electrostatic Vlasov-Poisson simulation, the application to the electromagnetic Vlasov simulation of magnetized plasma is still limited, mainly owing to the difficulty in solving the gyro motion around the magnetic field line.

In this paper, we propose a new numerical scheme for the advection equation, specifically designed to solve the Vlasov equation in magnetized plasma. The scheme is briefly introduced in Section \ref{sec:multi-moment-advect}. Benchmark tests of the scheme and its application to the Vlasov simulation are presented in Section \ref{sec:numerical-tests}. Finally, we summarize the paper in Section \ref{sec:summary}. Details of the scheme have been presented in \cite{2011JCoPh.230.6800M}.

\section{Multi-Moment Advection scheme}\label{sec:multi-moment-advect}
The present scheme considers the advection of a profile $f(\vect{x},t)$ and its zeroth to second order moments defined as
\begin{eqnarray}
\vect{M}^{m} = \frac{1}{m!}\inty{}{}\vect{x}^{m} f d\vect{x}, \;\;\; \left(m=0,1,2\right).\label{eq:1} 
\end{eqnarray}
In one dimension, their governing equations are written as
\begin{eqnarray}
&& \pdif{f}{t} + \pdif{}{x}\left(u f\right) = 0,\label{eq:2}\\
&& \pdif{M^0}{t} + \inty{}{}dx \pdif{}{x}\left(u f\right) = 0,\label{eq:3}\\
&& \pdif{M^m}{t} + \frac{1}{m!} \inty{}{}dx \pdif{}{x}\left(u x^{m} f\right)= \frac{1}{\left(m-1\right)!}\inty{}{}ux^{m-1}fdx,\;\;\; \left(m=1,2\right),\label{eq:4}
\end{eqnarray}
where $u$ is the velocity. Equations (\ref{eq:3}) and (\ref{eq:4}) are exactly obtained by multiplying Equation (\ref{eq:2}) by $x^{m}/m!$ and then integrating over space.
To solve a set of these equations, the one-dimensional scheme treats four dependent variables; the point value of the profile $f_i$, and the piecewise moments,
\begin{eqnarray}
M^m_{i+1/2} = \frac{1}{m!}\inty{x_{i}}{x_{i+1}}x^{m}fdx,\;\;\; \left(m=0,1,2\right),\label{eq:5}
\end{eqnarray}
and constructs a piecewise interpolation for $f$ in a cell with a fourth order polynomial, $F_{i}(x) = \sum_{k=1}^{5} k C_{k;i} (x-x_i)^{k-1}.$
The five coefficients $C_{k;i}$ are explicitly determined from the dependent variables at the upwind position as constraint. Then the variables are advanced on the basis of their governing equations (\ref{eq:2})-(\ref{eq:4}) with the semi-Lagrangian method.

The two-dimensional scheme is designed in a similar way. It treats six dependent variables; the point value of the profile $f_{i,j}$, and the piecewise moments in the $x$ and $y$ directions,
\begin{eqnarray}
\vect{M}^m_{i+1/2,j+1/2} &=& \frac{1}{m!} \inty{y_{j}}{y_{j+1}} \!\!\! \inty{x_{i}}{x_{i+1}} \vect{x}^m fdxdy, \;\;\; \left(m=0,1,2\right),\label{eq:7}
\end{eqnarray}
and constructs a piecewise interpolation for $f$ in a cell with a quadratic polynomial, $F_{i,j}(x,y) = \sum_{l=1}^{3} \sum_{k=1}^{3} lk C_{lk;i,j} (x-x_{i})^{k-1} (y-y_{j})^{l-1}.$
The nine coefficients $C_{lk;i,j}$ are explicitly determined from the dependent variables at the upwind position as constraint.
The scheme is termed as the ``Multi-Moment Advection (MMA)'' scheme.
For details, see \cite{2011JCoPh.230.6800M}. 

\section{Benchmark tests}\label{sec:numerical-tests}

\begin{figure}[t]
\centering
\includegraphics[clip,angle=0,scale=.32]{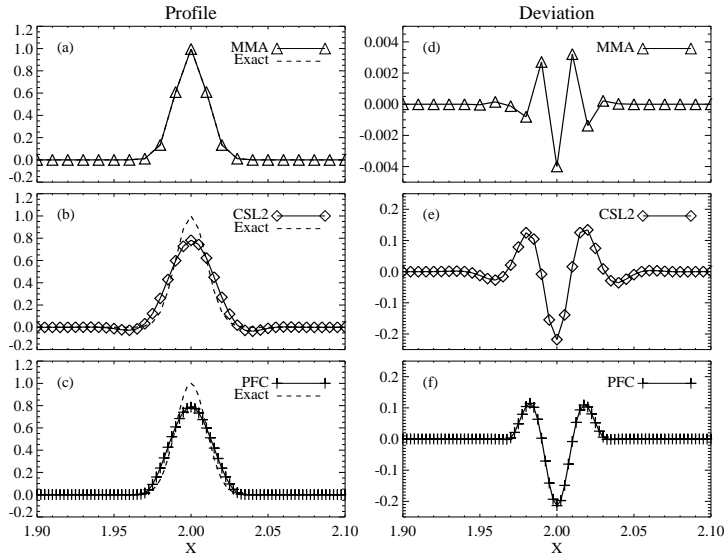}
\caption{One-dimensional linear advection of a gaussian profile. (a,b,c) Calculation results with the MMA, the CIP-CSL2, and the PFC schemes (solid lines with symbols). Dashed lines are the exact solution. (d,e,f) Deviation of the calculation results from the exact solution.}
\label{fig:adv_1d}
\end{figure}

Figure \ref{fig:adv_1d}(a-c) shows the one-dimensional linear advection problem of a gaussian profile solved by the MMA, CIP-CSL2 \citep{2001mwr...129..332Y}, and PFC \citep{2001JCoPh.172..166F} schemes. Since the numbers of dependent variables are different among the three schemes (four for the MMA, two for the CIP-CSL2, and one for the PFC), we use different grid sizes so that the total memory usage is equal. The CFL number is 0.2. 
The MMA scheme (a) provides a quite accurate solution compared to other schemes (b,c).
{Figure \ref{fig:adv_1d}(d-f) shows the deviation of the calculation results from the exact solution. The MMA scheme (d) is about fifty times better then other schemes (e,f).}

\begin{figure}[t]
\centering
\includegraphics[clip,angle=0,scale=.36]{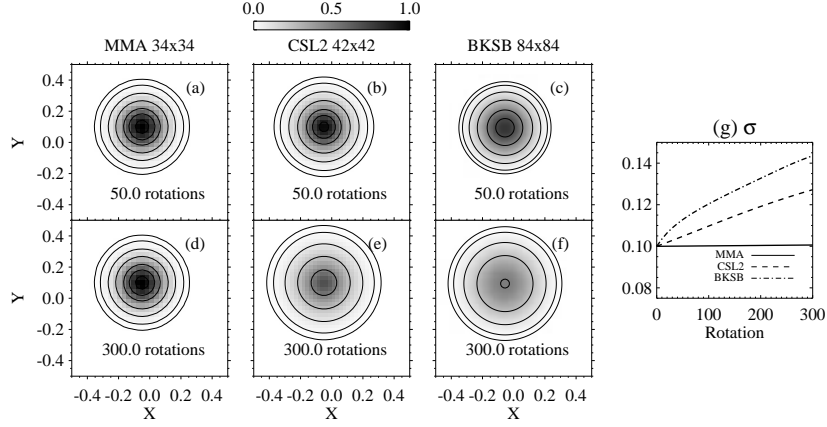}
\caption{Two-dimensional solid body rotation and advection of a symmetric gaussian profile. (a,b,c) Calculation results after 50 rotations with the MMA, the CIP-CSL2, and the backsubstitution schemes. (d,e,f) Calculation results after 300 rotations. (g) Temporal variation of the standard deviation $\sigma$. Solid, dashed, and dot-dashed lines are obtained from the MMA, CIP-CSL2, and backsubstitution schemes.}
\label{fig:adv_rot_2d}
\end{figure}

Figure \ref{fig:adv_rot_2d}(a-f) shows the two-dimensional solid body rotation and advection problem of a symmetric gaussian profile,
\begin{eqnarray}
\frac{\partial f}{\partial t}-\left(y-y_0\right)\pdif{f}{x}+\left(x-x_0\right)\pdif{f}{y}=0,\;\;\;
f\left(x,y,t=0\right) = \exp\left[-\frac{\left(x-x_0\right)^2+\left(y-y_0\right)^2}{2 \sigma^2}\right],\nonumber
\end{eqnarray}
solved by the MMA, CIP-CSL2 \citep{2002CoPhC.148..137T}, and backsubstitution \citep{2006CoPhC.175...86S} schemes. This describes the rotation around $(x,y)=(x_0,y_0)$, corresponding to the electric field drift motion for magnetized plasma. The parameters are $(x_0,y_0,\sigma) = (-0.05,0.1,0.1)$, and the simulation domain is $[-0.5,0.5]$ in both directions.
Since the numbers of dependent variables are different among the three schemes (six for the MMA, four for the CIP-CSL2, and one for the backsubstitution), we use the different numbers of grid points ($34\times34$ for the MMA, $42\times42$ for the CIP-CSL2, and $84\times84$ for the backsubstitution) so that the total memory usage is equal.
The time steps are 0.004$\pi$ for the MMA and CIP-CSL2 schemes, and 0.002$\pi$ for the backsubstitution scheme so that the CFL number is close among the three simulations.
While other schemes show serious numerical diffusion after several tens of rotation periods, the MMA scheme completely preserves the profile after hundreds of rotation periods.
{Figure \ref{fig:adv_rot_2d}(g) shows the temporal variation of the standard deviation $\sigma$ obtained by fitting the profile with the gaussian function.
After 300 rotation periods, the standard deviation is increased by 0.1006 (MMA), 0.1272 (CIP-CSL2), and 0.1436 (backsubstitution).}

We apply the MMA scheme to the electromagnetic Vlasov-Maxwell simulation. The one-dimensional Vlasov-Maxwell system of equations is written as
\begin{eqnarray}
&& \pdif{f_s}{t}+v_x\pdif{f_s}{x}+\frac{q_s}{m_s}\left(\vect{E}+\frac{\vect{v}\times\vect{B}}{c}\right)\cdot\pdif{f_s}{\vect{v}} = 0,\;\;\;\left(s=p,e\right),\label{eq:10}\\
&& \pdif{\vect{E}}{t}=c \nabla \times \vect{B} - 4 \pi \vect{j},\;\;\;
 \pdif{\vect{B}}{t}=-c \nabla \times \vect{E},\;\;\;
 \vect{j} = \sum_{s=p,e}q_s \inty{}{}\vect{v}f_s d\vect{v},\label{eq:11}
\end{eqnarray}
where $\vect{E}$ and $\vect{B}$ are the electric and magnetic fields, $\vect{j}$ is the current density, $c$ is the speed of light, $q_s$ is the charge, $m_s$ is the mass, and the subscript $s$ denotes particle species ($p$ for protons and $e$ for electrons). We assume the two dimensionality in velocity space, $\vect{v} = (v_x,v_y,0)$, $\vect{E} = (E_x,E_y,0)$, and $\vect{B} = (0,0,B_z)$. The Vlasov equation (\ref{eq:10}) is split into two equations in two-dimensional velocity and one-dimensional configuration spaces, which are advanced by the MMA and CIP-CSL2 schemes, respectively. The Maxwell equation (\ref{eq:11}) is solved by the CIP scheme \citep{2006CiCP.1..311O}. The time integration of the system is carried out in the same manner as \cite{2011JCoPh.230.6800M}.

\begin{figure}[t]
\centering
\includegraphics[clip,angle=0,scale=.35]{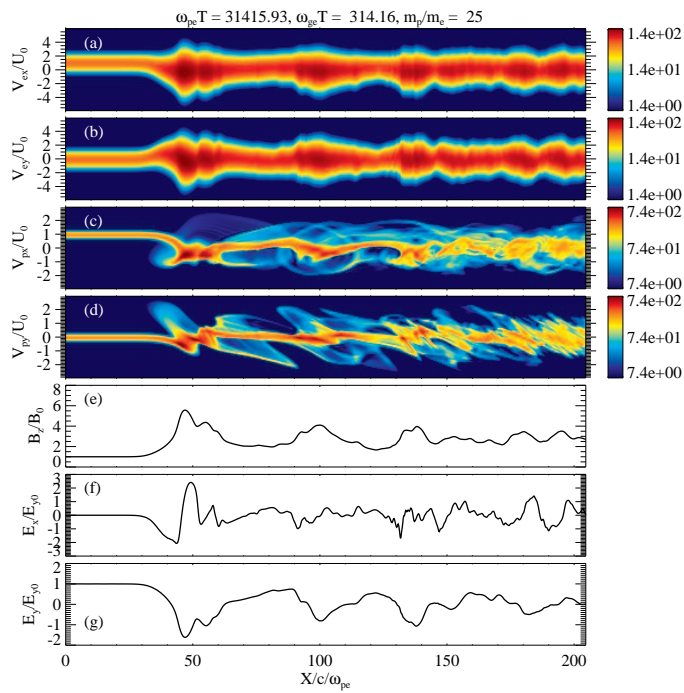}
\caption{One-dimensional electromagnetic Vlasov simulation of perpendicular shock waves. (a,b) The electron phase space distributions in $(x,v_{ex})$ and $(x,v_{ey})$. (c,d) The proton phase space distributions in $(x,v_{px})$ and $(x,v_{py})$. (e,f,g) The electromagnetic field $(B_{z},E_{x},E_{y})$ distributions. The velocity and the electric field are normalized by the bulk flow velocity and the motional electric field at the upstream, respectively.}
\label{fig:shock}
\end{figure}

We simulate one-dimensional, strictly perpendicular collisionless shock waves \citep[e.g.,][]{2002ApJ...572..880H}. A high speed plasma is injected from the left-hand boundary and flows toward positive direction. The plasma carries the perpendicular magnetic field. At the right-hand boundary, the plasma is specularly reflected. As a result, a shock wave is formed and propagates in negative direction. 
Simulation parameters are as follows; a proton to electron mass ratio $m_p/m_e = 25$, a ratio of the electron plasma to gyro frequency $\omega_{pe}/\omega_{ge} = 100$, electron and proton plasma beta values $\beta_{e} = \beta_{p} = 1.0$, and an {\Alfven} Mach number of the upstream plasma flow is $5.0$. The velocity space domain is $[-0.06c,0.06c]$ for electrons and $[-0.03c,0.03c]$ for protons with 72 grid points in both the $v_x$ and $v_y$ directions. 
The configuration space domain is $20480 \lambda_{D}$ with 1024 grid points $(\Delta x = 20 \lambda_{D})$, where $\lambda_{D}$ is the Debye length.
The time step is $\Delta t = 0.1\pi \omega_{pe}^{-1}$. 

Figure \ref{fig:shock} shows the electron phase space distribution $(\inty{}{}f_{e}dv_{y}, \inty{}{}f_{e}dv_{x})$, the proton phase space distribution $(\inty{}{}f_{p}dv_{y}$,$\inty{}{}f_{p}dv_{x})$, and the electromagnetic fields $(B_{z},E_{x},E_{y})$ at $\omega_{ge} t=100 \pi$. 
An {\Alfven} Mach number of the resulting shock wave is $\sim 7.5$ measured in the shock rest frame.
The simulation describes fundamental structures of the perpendicular collisionless shock.
The plasma pressure and the magnetic field strength rise at the shock front $(x=50)$, and subsequently oscillate due to the gyro motion of protons in the downstream region $(x > 50)$. 
Around the front, the difference in inertia between electrons and protons produces the electrostatic potential (Figure \ref{fig:shock}(f), the so-called shock potential).
Before the front, there is a gradual increase of the magnetic field strength ($30 < x < 40$), in which part of protons are reflected by the shock potential (Figure \ref{fig:shock}(c), the so-called reflected ions).
 We confirm that the Rankine-Hugoniot conservation laws are satisfied at the shock.
 Even at this moment, the electron magnetic moment is well conserved in the downstream region, due to the fact that the MMA scheme can solve the solid body rotation with little numerical diffusion.

\section{Summary}\label{sec:summary}
We have presented a new numerical scheme for solving the advection equation and the Vlasov equation.
The present scheme solves not only point values of a profile but also its zeroth to second order piecewise moments as dependent variables, and advances them on the basis of their governing equations.
We have developed one- and two-dimensional schemes, and have shown their high capabilities.
The scheme provides quite accurate solutions compared to other existing schemes with the same memory usage. 
The two-dimensional scheme can solve the solid body rotation problem of a gaussian profile with little numerical diffusion.
This is a very important property for Vlasov simulations of magnetized plasma.

The application of the scheme to the electromagnetic Vlasov simulation of collisionless shock waves has been presented. In the simulation, we use $\omega_{pe}/\omega_{ge} = 100$ and $\Delta x = 20 \lambda_{D}$. 
Although the grid size is much larger than the Debye length so that the Debye-scale structures can not be described, the simulation is stable and the meso-scale shock structures are well described. 
Since the grid size of an explicit PIC simulation is restricted to the Debye length, the PIC simulation requires large computational cost when $\omega_{pe}/\omega_{ge}$ is large.
This is not the case for Vlasov simulations, unless Debye-scale structures are important.
This advantage enables us to perform large-scale plasma kinetic simulations with large $\omega_{pe}/\omega_{ge}$.

\acknowledgements 
We thank an anonymous referee for reviewing the manuscript.
Part of Figures \ref{fig:adv_1d} and \ref{fig:adv_rot_2d} are reproduced by permission of the Elsevier Inc.


\end{document}